\documentclass[traditabstract]{aa}
\usepackage{natbib}
\usepackage{amssymb}
\usepackage{graphicx}
\usepackage{epsfig}

\begin{document}
\title{The influence of electron collisions on non-LTE Li line formation in stellar atmospheres}
\author{Y. Osorio \inst{1}
	 \and  P. S. Barklem \inst{1} \and K. Lind \inst{2} \and M. Asplund\inst{2}}
\institute{Department of Physics and Astronomy, Uppsala University, Box 516, 751 20 Uppsala, Sweden
	\and Max-Planck-Institut f{\"u}r Astrophysik, Karl-Schwarzschild-Strasse 1, 857 41 Garching bei M{\"u}nchen, Germany}
\offprints{Yeisson Osorio}
\mail{yeisson.osorio@fysast.uu.se}


   \date{Received 29 Dec 2010 ; accepted 1 Feb 2011}

  \abstract
   {The influence of the uncertainties in the rate coefficient data for electron-impact excitation and ionization on non-LTE Li line formation in cool stellar atmospheres is investigated.  We examine the electron collision data used in previous non-LTE calculations and compare them to recent calculations that use convergent close-coupling (CCC) techniques and to our own calculations using the R-matrix with pseudostates (RMPS) method.  We find excellent agreement between rate coefficients from the CCC and RMPS calculations, and reasonable agreement between these data and the semi-empirical data used in non-LTE calculations up to now.  The results of non-LTE calculations using the old and new data sets are compared and only small differences found: about 0.01 dex ($\sim 2$\%) or less in the abundance corrections.   We therefore conclude that the influence on non-LTE calculations of uncertainties in the electron collision data  is negligible.  Indeed, together with the collision data for the charge exchange process $\mathrm{Li(3s)} + \mathrm{H} \rightleftharpoons \mathrm{Li}^+ + \mathrm{H}^-$ now available, and barring the existence of an unknown important collisional process, the collisional data in general is not a source of significant uncertainty in non-LTE Li line formation calculations.}

   \maketitle

\section{Introduction}

Lithium abundances in stellar atmospheres are key observables in astrophysics, giving crucial information on stellar evolution and mixing, stellar and Big Bang nucleosynthesis, cosmic ray spallation, and perhaps even planet formation \citep[e.g.][]{1993PhST...47..186L, 1994A&A...288..860C, 2010Ap&SS.328..193M}.  Such abundances are interpreted from observations of the few Li I lines found in stellar spectra, and to obtain accurate results, it is important to account for departures from local thermodynamic equilibrium (LTE) \citep{1984A&A...130..319S,1994A&A...288..860C, Lind2009}.  Reliable modelling of line formation in non-LTE requires detailed knowledge of all important radiative and collisional processes on the atom of interest, Li.  

The collisional processes are particularly challenging. For many years following the pioneering study of \citet{1984A&A...130..319S}, the importance of inelastic hydrogen collisions was a major uncertainty.   Detailed quantum scattering calculations of Li+H collisions \citep{H+Li} and application to non-LTE modelling \citep{2003A&A...409L...1B,Lind2009} have allowed this question to be answered, and it was found that direct excitations by hydrogen collisions essentially have no influence.  However, a related charge-exchange process $\mathrm{Li(3s)} + \mathrm{H} \rightleftharpoons \mathrm{Li}^+ + \mathrm{H}^-$ was shown to be important, resulting in differences in derived abundances of about 0.05 dex in solar-metallicity stars and 0.1 dex in metal-poor stars.  Moreover, it was found that the results were not sensitive to uncertainties in the data for this process.   When the rate coefficient was altered by factors in keeping with the expected uncertainty in the theoretical calculation, the effects on the line formation were practically negligible ($< 0.01$ dex in derived abundances).  

Thus, barring there being additional important collisional processes on Li we are unaware of, the uncertainties regarding collisional data lie, perhaps, with the electron collisions.  The question of quantifying these has only been touched upon briefly in the past.  \citet{1994A&A...288..860C} made calculations exploring the sensitivity of their results to various input parameters, including collision cross sections.  They identified the oscillator strengths of the lines of interest (6708 and 6104 \AA) and the photoionization cross sections from $2s$ and $2p$ as the most important atomic data in determining the uncertainties of their calculations. They estimate an error in the abundance corrections of less than 0.01 dex arising from these sources.  That the sensitivity to collisional cross sections was explored, though not discussed explicitly implies even smaller uncertainties due to collision cross sections, but details have not been published.  This study and others \citep[e.g.][]{1984A&A...130..319S, Lind2009} make use of empirically corrected calculations by \citet{1962ApJ...136..906V} and \citet{Park1971} for excitation and \citet{1976asqu.book.....A} or similar for ionization.  

Given the great astrophysical importance of Li abundances and the considerable advances in calculation methods for electron scattering in the intervening period, in particular advanced close-coupling methods that are able to account for effects of coupling to the target continuum, we considered it worthwhile and prudent to accurately quantify the uncertainties associated with the electron collision data.  In this paper we examine the electron collision data in the literature, and perform a new $R$-matrix calculation, in order to accurately estimate the uncertainties in such calculations.  The data are then used in non-LTE Li I line formation modelling to assess the resulting uncertainties in stellar Li abundances.

\section{Electron collision data}

In this section we examine the existing data for excitation and ionization by electron impact used in non-LTE calculations.  We compare with advanced close-coupling calculations including our own calculations for the excitation processes and present the details.

\subsection{Excitation}

There are a number of existing calculations for electron-impact excitation $\mathrm{Li}(nl) + e \rightarrow \mathrm{Li}(n^\prime l^\prime) + e$, though only relatively recently have there been any large-scale calculations including a significant number of excited states.  Two advanced close-coupling methods have been used, namely convergent close-coupling (CCC, \citealt{PhysRevA.46.6995}) and $R$-matrix with pseudostates (RMPS, \citealt{0953-4075-29-1-015}) approaches.  The advantage of the CCC method is that, at least for collision systems with two electrons, it has been shown to yield convergent results as basis size is increased; however, it has the disadvantage of requiring considerable computational effort.  The advantage of the RMPS method is that the $R$-matrix is calculated once and then results for many collision energies can be calculated with little additional effort.  This is particularly important for the study of resonances requiring dense energy grids.  The disadvantage, however, is that convergence with basis size can be slower resulting in the appearance of pseudo-resonances.

\citet{CCC} have published data from extensive 45-state CCC calculations for transitions with $n\le 3$ and $n^\prime \le 4$, and include semi-empirical cross sections for $n=4$, $n^\prime=4$.  The calculations show good agreement with experiments where available, and analytic fits to the cross section data are provided.  \cite{Griffin2001} have done calculations using a 55-state RMPS approach resulting in cross section data for all transitions with $n=2$, $l=0$ and $n^\prime \le$ 4.  Their calculations demonstrate the importance of including the coupling to the target continuum via pseudostates, particularly at collisions energies greater than the ionization energy, i.e. $> 5.4$~eV, where the effects are small for $2s\rightarrow 2p$, but become significant for $2s\rightarrow 3l$ and even greater for $2s\rightarrow 4l$.  The cross sections are in good agreement with the CCC and experimental results where compared.  Since there is a possibility that resonances at low collision energies are unresolved by the CCC calculations and could contribute significantly to the rate coefficients, we performed our own RMPS calculations for $n \le$ 4 and $n^\prime \le$ 4 and $n^\prime l^\prime = 5s$, which gives the additional advantage of an independent check.  These calculations are described below in Sect.~\ref{sect:rmat}  and compared with the existing data in Sect.~\ref{sect:res}.

As mentioned above, the rate coefficients calculated by \citet{Park1971} have been used in most calculations of non-LTE formation of Li I lines in cool stars \citep{1984A&A...130..319S,1994A&A...288..860C, Lind2009,2010A&A...522A..26S,2005PASJ...57...45T}, so they represent an important basis for comparison in this work.  \citet{1984A&A...130..319S} replaced the data of the important $2s\rightarrow 2p$ transition with that of \citet{1962ApJ...136..906V}.  Both these sources are based on Born approximation calculations, which are well known to be valid only at high impact energies and to substantially overestimate the cross sections near threshold.  Thus, empirical corrections determined from various experiments are applied.  Based on comparison with the experiments, both works conclude that the calculated rates are not likely to be in error by more than a factor of two.

\subsubsection{R-matrix calculation}
\label{sect:rmat}

We made a 34-state RMPS calculation of the excitation by electron impacts, using existing freely available computer codes.  These calculations will now be described.

For the initial atomic structure calculations we used the code \texttt{CIVPOL}  (\citealt{civpol_thesis,CIVPOL}; see also \citealt{2004JPhB...37.2979P}), which is an adaptation of \texttt{CIV3} \citep{CIV3} that allows the construction of polarization pseudostates.  Such pseudostates are used to account for coupling to the target continuum and are needed to obtain a correct description of the dipole polarization and thus the long-range interaction potential.  They have been shown by \citet{Griffin2001} to be important at intermediate energies.  \texttt{CIV3} and \texttt{CIVPOL} use the configuration interaction (CI) method and the radial parts of the (pseudo) orbitals are represented by Slater-type orbitals; for details see \mbox{\cite{CIV3}} and \cite{CIVPOL}.
The atomic structure of the target Li I was built by optimising spectroscopic orbitals with all allowed $nl$ values up to $nl=5f$; the $1s$ and $2s$ orbitals were Hartree-Fock orbitals taken from \mbox{\cite{clementi}}.   The energies obtained for the ten lowest-lying spectroscopic states are compared with experimental values taken from the NIST Atomic Spectra Database \citep{NIST} in Table \ref{energy}, and the agreement is satisfactory.

\begin{table}
\caption{Comparison with experiment for the energies of the lowest states of Li I.}\label{energy}
\centering
	\begin{tabular}{l l c c c}\hline
   \multicolumn{2}{c}{State}         &   \multicolumn{2}{c}{Energy (Ry)}  &   \\
                                &                      &   This Work &  Experiment & Difference \\\hline\hline\\[-0.2 cm]
$1s^2$ $2s$                      &  $^2$S     &          ---          &          ---        &        ---         \\   
\phantom{$1s^2$ }$2p$  & $^2$P$^\mathrm{o}$  &     0.1350      &     0.1358    &    $-$0.59\%    \\
\phantom{$1s^2$ }$3s$  & $^2$S       &     0.2474      &     0.2479    &    $-$0.20\%    \\
\phantom{$1s^2$ }$3p$  & $^2$P$^\mathrm{o}$  &     0.2807      &     0.2818    &    $-$0.39\%    \\
\phantom{$1s^2$ }$3d$  & $^2$D      &     0.2839      &     0.2851    &    $-$0.42\%    \\
\phantom{$1s^2$ }$4s$  & $^2$S       &     0.3180      &     0.3191    &   $-$0.34\%     \\
\phantom{$1s^2$ }$4p$  & $^2$P$^\mathrm{o}$  &     0.3312      &     0.3323     &   $-$0.33\%    \\
\phantom{$1s^2$ }$4d$  & $^2$D      &     0.3325      &     0.3337     &  $-$0.36\%     \\
\phantom{$1s^2$ }$4f$   & $^2$F$^\mathrm{o}$   &     0.3326      &     0.3338     &   $-$0.36\%    \\ 
\phantom{$1s^2$ }$5s$  & $^2$S       &     0.3479      &     0.3490     &   $-$0.32\%    \\\hline\hline
	\end{tabular}
	\tablefoot{Energies are relative to the ground state and compared with experimental values obtained from NIST data (see text). The energy difference is measured $(E_\mathrm{Theor}-E_\mathrm{Exper})/E_\mathrm{Exper}$.}
\end{table}

Once the spectroscopic orbitals (and states) were calculated, we added $n=6$($s,p,d$) and 7($s,p,d$) pseudo-orbitals to describe the polarization of the ground and first excited state, respectively. To test the quality of the polarization pseudostates, we compared the static dipole polarizabilities obtained with experimental and theoretical values found in the literature, see Table~\ref{polar}. For the polarization of the ground state there is good agreement between our results and those from the literature. In the case of the polarizability of the first excited state, we found only one other theoretical result \citep{PolThe-a}, where the polarizability is calculated in the Coulomb approximation. 

\begin{table}
\caption{Dipole polarizabilities of the ground and first excited state compared with values found in the literature. }\label{polar}
\centering
	\begin{tabular}{c  c  l}\hline
	              State                 &    \multicolumn{2}{c}{Polarizability ($a_0^3$)}\\
	                                        &      This Work    &   \multicolumn{1}{c}{Other Works} \\\hline\hline
	$1s^2\,2s$ ($^2$S)       &        168.5          &   (a) 162.3, (b) {\bf 164.0(34)}    \\
                                                 &                             &   (c) 164.1 \\\hline
	$ 1s^2\,2p$ ($^2$P$^\mathrm{e}$) &        148.5           &   (a) 117.8     \\\hline\hline
	\end{tabular}
	\tablebib{(a) \cite{PolThe-a} using Coulomb Approximation; (b) \cite{PolExp} using E-H gradient balance technique; (c) \cite{PolThe-c} using CI-Hylleraas. Bold values are experimental.}
\end{table}
 
The computed orbitals were then used in a RMPS calculation of the electron-impact excitation cross sections.  The {\tt RMATRX I} code \citep{InnerRmatrix} was used for the internal region problem, and for the external region problem we employed a version of the code {\tt STGF} \citep{OuterRmatrix} modified for treating collisions with neutral atoms \citep{Badnell}.  
 
After exploration of the dependence of our results on the calculation details, we finally adopted the following.  The target atom consists of 30 spectroscopic states (the ten states in Table \ref{energy} with an additional 20 spectroscopic states) and four polarization pseudostates; 34 states in total.  Target atoms with 40 and 45 states were tested and showed no significant change in the cross section results.  For both the $N$-electron and $N+1$-electron configurations, maximum occupation numbers of 2 for the spectroscopic orbitals and 1 for the pseudo-orbitals were adopted, with double excitations permitted. Calculations extending the $N+1$-electron configurations to allow all possible ways of adding one electron to the $N$-electron configurations showed no significant differences in the inelastic cross sections.  The $R$-matrix boundary between the inner and outer regions was set at 83.4 a.u.  Partial waves up to $L=50$ were calculated.

\subsubsection{Results and comparison}
\label{sect:res}

\begin{figure*}[t]
\centering
\includegraphics[width=0.95\textwidth]{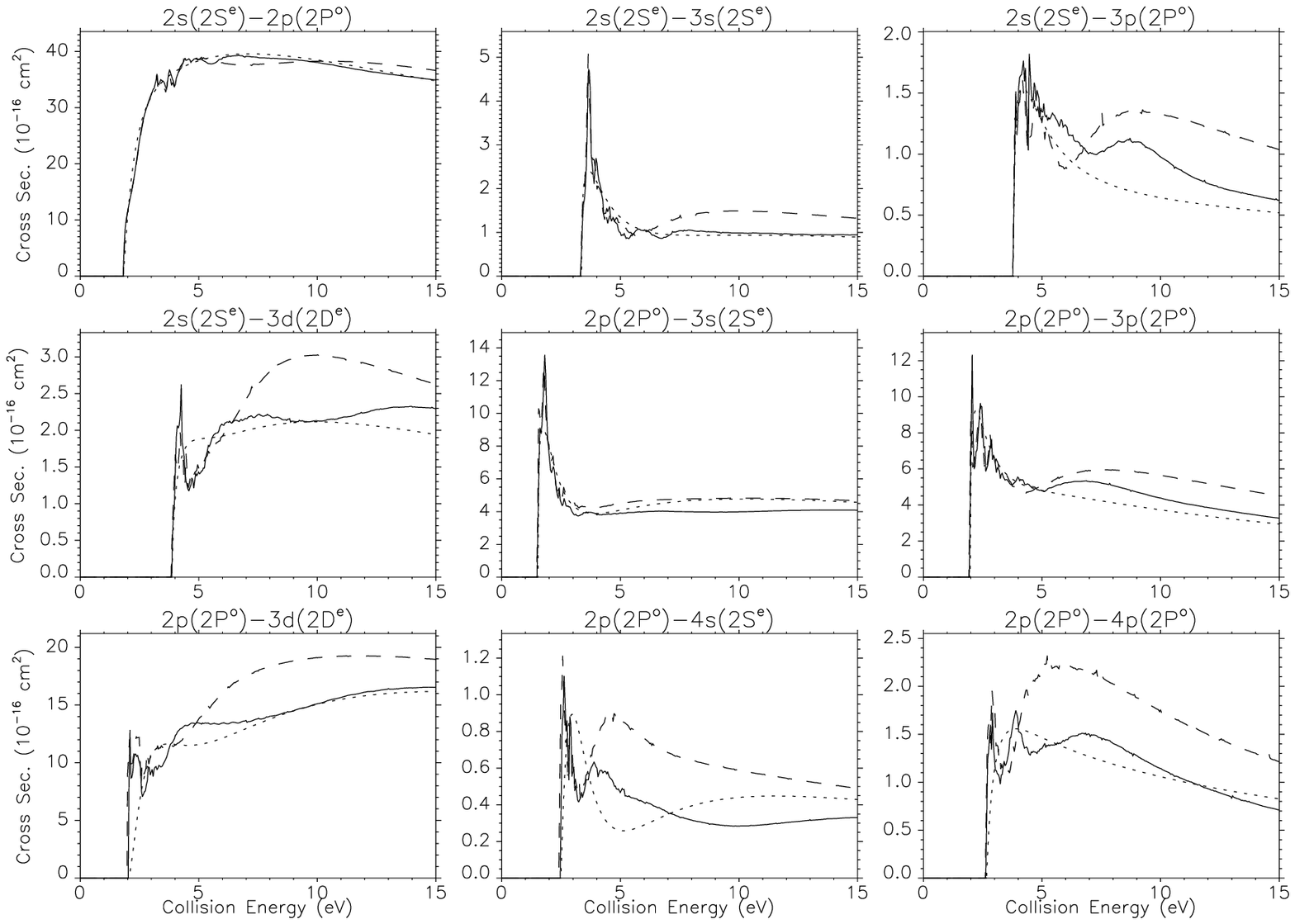}
     \caption{Comparison of electron-impact excitation cross sections with collision energy  for some transitions involving the lowest states of Li I.  Solid lines show our calculation using the polarization pseudo-orbitals, dashed lines are the cross sections obtained without using pseudo-orbitals, and dotted lines are the fits to the CCC calculations \citep{CCC}.}\label{crossec}
\end{figure*}

In Fig.~\ref{crossec} some of the resulting cross sections are shown, and then compared with a calculation without pseudostates, and with the fits to the CCC results of \citet{CCC}.  It is seen that the final cross sections compare well with the CCC results, particularly at low collision energies, of greatest interest for the rates at cool star temperatures.  It appears that our results are not always fully converged at intermediate energies.  For example, at collision energies above 5.4 eV, where processes involving the continuum are directly accessible, the $2s \rightarrow 3p$ cross section lies between the CCC result and the calculation without pseudostates.  \cite{Griffin2001} obtain good agreement between their somewhat larger 55-state RMPS calculation and the CCC result.  Thus, we suspect that the intermediate-energy results are not converged in this and in some other transitions; we note that this may be a result of the smaller basis size, but could also come from other details of the calculation.  However, as we will see, this is not important for calculating the rate coefficients at the temperatures and precision of interest for stellar atmosphere applications.  The behaviour within a couple of eV of threshold is far more important.  Examining Fig.~\ref{crossec} we find generally good agreement between the CCC results and ours in this region.

\begin{figure*}
\centering
\begin{tabular}{c c c}
\includegraphics[width=0.31\textwidth]{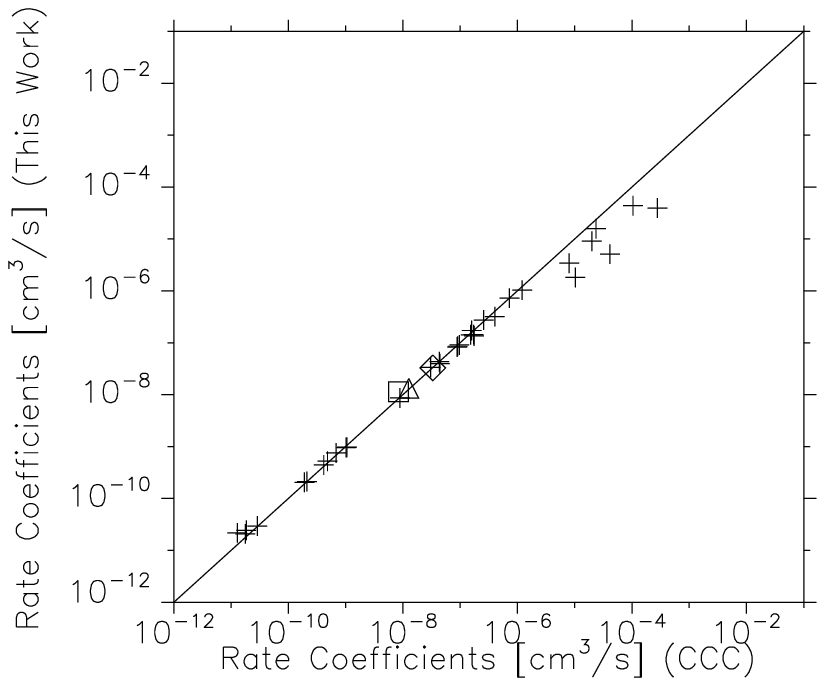}&\includegraphics[width=0.31\textwidth]{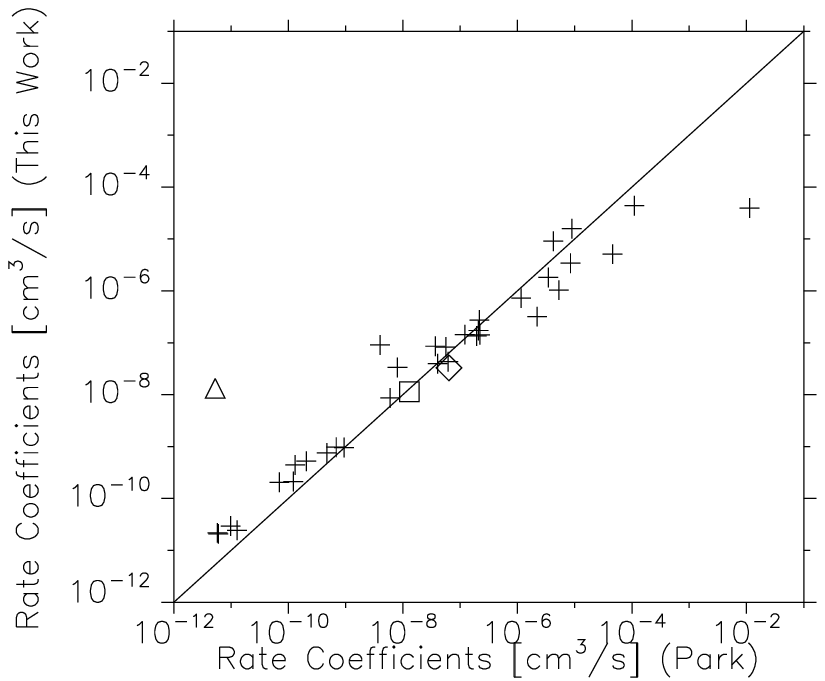}&\includegraphics[width=0.31\textwidth]{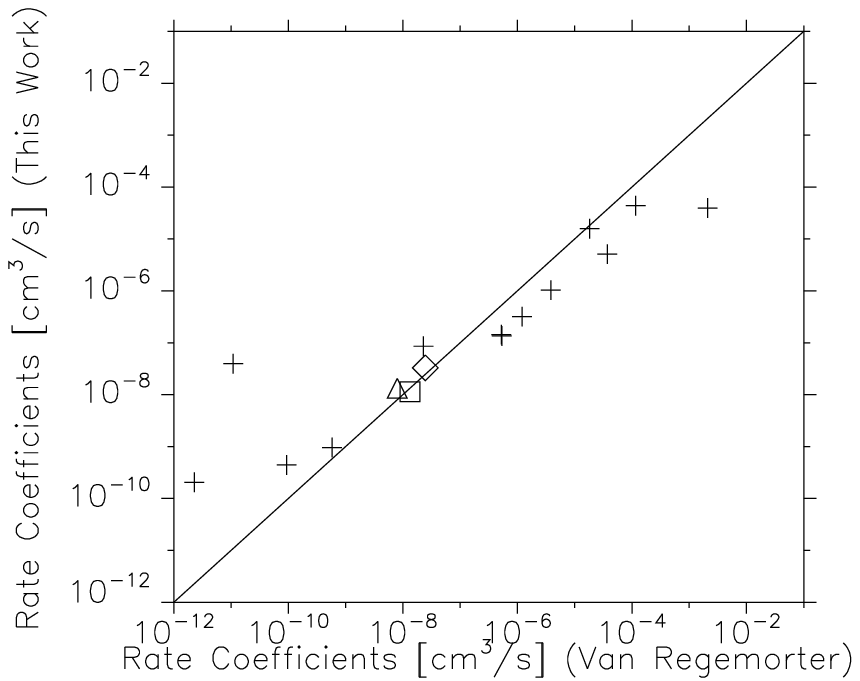}\\
\end{tabular}
\caption{Rate coefficients at 8000 K obtained by us compared with those calculated from the CCC calculations \cite[left]{CCC}, those of \cite[middle]{Park1971} and those calculated using the van Regemorter formula \citep{1962ApJ...136..906V}.  The latter is only valid for optically allowed transitions, hence the limited number of comparison points. The different symbols are for the rate coefficients corresponding to the transitions: \mbox{$2s \rightarrow 2p$ [ {\Large $\diamond$}, 6708 \AA\ ]}, \mbox{$2p \rightarrow 3d$ [ $\square$, 6104 \AA\ ]}, and \mbox{$2p \rightarrow 3s$ [ $\Delta$, 8126 \AA\ ]}. }\label{rates}
\end{figure*}

Rate coefficients were computed both from our RMPS cross sections and the CCC cross sections from \citet{CCC} by integrating the cross sections folded with a Maxwellian velocity distribution.  The rate coefficients from our RMPS calculations are presented in Table~\ref{tablerates} for a range of temperatures.  In the table the excitation rate coefficients are presented and de-excitation rates may be obtained from the detailed balance relation.    The two sets of rates are compared in the leftmost panel of Fig.~\ref{rates}, and the agreement is excellent.  This suggests that neither the inclusion of resonances nor the proper convergence at intermediate energies is vital at temperatures of interest, though we note that some cancellation may occur.  Assuming a log-normal distribution for the ratio $\langle\sigma v\rangle_\mathrm{this\,\, work}/\langle\sigma v\rangle_\mathrm{CCC}$ we find a mean of 0.83 and a scatter of 1.86; i.e. an offset of less than 20\% and a scatter of less than a factor of 2.  There is a tendency for our rate coefficients to be lower than the CCC ones for the transitions with the highest rate coefficients.  Of these seven transitions grouped with rate coefficients greater than $10^{-6}$~cm$^3$~s$^{-1}$, the one lying roughly on the one-to-one relation is $3p \rightarrow 3d$.  The remaining six transitions correspond to the $n = 4$ and $n^\prime = 4$ transitions, where, as mentioned earlier, the \citet{CCC} data are not actually CCC calculations, but semi-empirical data.  The agreement for the most important low-lying transitions (shown as different symbols) is excellent.   

Our data are also compared with the data from \cite{Park1971}  and data calculated using the van Regemorter formula \citep{1962ApJ...136..906V}, the comparisons shown in the middle and rightmost panels of Fig.~\ref{rates}, respectively.   Comparison with the data of Park is generally good, although the scatter is significantly larger than seen for the CCC rates.  The Park values are typically around 40\% greater than ours, and the scatter is around a factor of 6.  It is also notable that the rate for $2p\rightarrow 3s$, corresponding to the 8126~{\AA} line, is more than 3 orders of magnitude greater in our calculations than found by Park.  The $4d \rightarrow 4f$ is the other significantly discrepant transition. The van Regemorter formula is only valid for optically allowed transitions, showing a smaller number of comparisons.  The comparison with van Regemorter has an even greater scatter than found for Park, around a factor of 18, though owing to the small number of comparisons the scatter is dominated by a small number of outliers.   The offset is slightly smaller than for the Park data.  Van Regemorter's formula gives a rate for the $2p \rightarrow 3s$ transition in much better agreement with the rate from our calculation and the CCC data, while a few other transitions show large discrepancies.

\begin{table*}
\caption{Rate coefficients for inelastic collisions between electrons and Li I atoms for different temperatures.}\label{tablerates}
\centering
{\scriptsize
	\begin{tabular}{l c c c c c c c c c}\hline
Initital   &  \multicolumn{9}{c}{Final state}\\
  state  &    2p ($^2$P$^\mathrm{o}$)     &     3s ($^2$S$^\mathrm{e}$)    &     3p ($^2$P$^\mathrm{o}$)    &  3d ($^2$D$^\mathrm{e}$)  &  4s ($^2$S$^\mathrm{e}$)     &    4p ($^2$P$^\mathrm{o}$)    &    4d ($^2$D$^\mathrm{e}$)     &      4f ($^2$F$^\mathrm{o}$)    &            5s ($^2$S$^\mathrm{e}$) \\\hline\hline \\
     & \multicolumn{9}{c}{\bf 1 000 K}  \\
 2s ($^2$S$^\mathrm{e}$)   & 2.04($-$16)  & 1.93($-$24)  & 7.55($-$27)  & 3.51($-$27)  & 1.74($-$29)  & 6.23($-$31)  & 4.79($-$31)  & 4.76($-$31)  & 3.18($-$32)  \\
 2p ($^2$P$^\mathrm{o}$)   & - -  & 8.91($-$15)  & 5.30($-$17)  & 4.72($-$17)  & 2.81($-$20)  & 4.34($-$21)  & 3.17($-$21)  & 3.08($-$21)  & 7.00($-$23) \\
 3s ($^2$S$^\mathrm{e}$)  &   & - -  & 6.34($-$09)  & 3.94($-$09)  & 7.58($-$12)  & 7.02($-$13)  & 5.47($-$13)  & 7.16($-$13)  & 2.32($-$14)\\
 3p ($^2$P$^\mathrm{o}$)   &   &   & - -  & 1.12($-$05)  & 1.42($-$09)  & 2.07($-$10)  & 1.47($-$10)  & 1.63($-$10)  & 2.05($-$12)\\
 3d ($^2$D$^\mathrm{e}$)   &   &   &   & - -  & 1.62($-$09)  & 2.59($-$10)  & 1.96($-$10)  & 2.76($-$10)  & 3.10($-$12)\\
 4s ($^2$S$^\mathrm{e}$)   &   &   &   &   & - -  & 2.14($-$07)  & 1.88($-$07)  & 2.91($-$07)  & 5.41($-$09)\\
 4p ($^2$P$^\mathrm{o}$)   &   &   &   &   &   & - -  & 3.79($-$05)  & 1.33($-$05)  & 3.58($-$08)\\
 4d ($^2$D$^\mathrm{e}$)   &   &   &   &   &   &   & - -  & 6.11($-$05)  & 4.47($-$08)\\
 4f ($^2$F$^\mathrm{o}$)   &   &   &   &   &   &   &   & - -  & 4.57($-$08)\\
 
   &  \multicolumn{9}{c}{\bf 2 000 K}  \\
2s ($^2$S$^\mathrm{e}$) &    8.54($-$12) & 4.78($-$16) & 2.17($-$17) & 1.57($-$17) & 7.14($-$19) & 9.99($-$20) & 9.18($-$20) & 8.61($-$20) & 1.95($-$20) \\
2p ($^2$P$^\mathrm{o}$) & - -   &    4.17($-$11) & 3.13($-$12) & 3.61($-$12) & 2.70($-$14) & 1.49($-$14) & 1.32($-$14) & 1.22($-$14) & 9.37($-$16) \\
3s ($^2$S$^\mathrm{e}$) &    &  - -  &    8.29($-$08) & 7.68($-$08) & 1.40($-$09) & 3.92($-$10) & 3.84($-$10) & 5.52($-$10) & 5.12($-$11) \\
3p ($^2$P$^\mathrm{o}$) &    &    &  - -  &    1.67($-$05) & 1.96($-$08) & 9.20($-$09) & 7.71($-$09) & 9.31($-$09) & 3.56($-$10) \\
3d ($^2$D$^\mathrm{e}$) &    &    &    & - -   &    1.96($-$08) & 8.70($-$09) & 9.01($-$09) & 1.29($-$08) & 4.56($-$10) \\
4s ($^2$S$^\mathrm{e}$) &    &    &    &    & - -   &    7.34($-$07) & 8.27($-$07) & 9.39($-$07) & 4.94($-$08) \\
4p ($^2$P$^\mathrm{o}$) &    &    &    &    &    & - -   &    5.13($-$05) & 1.40($-$05) & 1.54($-$07) \\
4d ($^2$D$^\mathrm{e}$) &    &    &    &    &    &    & - -   &    5.99($-$05) & 1.61($-$07) \\
4f ($^2$F$^\mathrm{o}$) &    &    &    &    &    &    &    & - -   &    1.40($-$07) \\

   &  \multicolumn{9}{c}{\bf 5 000 K}  \\
2s ($^2$S$^\mathrm{e}$) &    5.91($-$09) & 3.88($-$11) & 8.32($-$12) & 8.06($-$12) & 1.16($-$12) & 4.52($-$13) & 5.23($-$13) & 4.24($-$13) & 2.25($-$13) \\
2p ($^2$P$^\mathrm{o}$) & - -   &    4.97($-$09) & 1.91($-$09) & 2.62($-$09) & 7.82($-$11) & 1.00($-$10) & 1.18($-$10) & 1.06($-$10) & 1.88($-$11) \\
3s ($^2$S$^\mathrm{e}$) &    & - -   &    4.96($-$07) & 4.75($-$07) & 2.32($-$08) & 1.50($-$08) & 1.53($-$08) & 3.14($-$08) & 4.88($-$09) \\
3p ($^2$P$^\mathrm{o}$) &    &    &  - -  &    1.96($-$05) & 9.03($-$08) & 8.17($-$08) & 8.09($-$08) & 1.24($-$07) & 8.32($-$09) \\
3d ($^2$D$^\mathrm{e}$) &    &    &    & - -   &    7.26($-$08) & 5.63($-$08) & 9.24($-$08) & 1.46($-$07) & 8.44($-$09) \\
4s ($^2$S$^\mathrm{e}$) &    &    &    &    & - -   &    2.71($-$06) & 2.47($-$06) & 1.65($-$06) & 1.74($-$07) \\
4p ($^2$P$^\mathrm{o}$) &    &    &    &    &    & - -   &    5.32($-$05) & 1.14($-$05) & 5.00($-$07) \\
4d ($^2$D$^\mathrm{e}$) &    &    &    &    &    &    & - -   &    4.70($-$05) & 3.54($-$07) \\
4f ($^2$F$^\mathrm{o}$) &    &    &    &    &    &    &    & - -   &    2.20($-$07) \\

  &  \multicolumn{9}{c}{\bf 8 000 K}  \\
2s ($^2$S$^\mathrm{e}$) &    3.25($-$08) & 5.44($-$10) & 1.86($-$10) & 2.07($-$10) & 3.71($-$11) & 2.05($-$11) & 2.80($-$11) & 2.02($-$11) & 1.25($-$11) \\
2p ($^2$P$^\mathrm{o}$) &  - -  &    1.43($-$08) & 8.63($-$09) & 1.32($-$08) & 5.39($-$10) & 8.85($-$10) & 1.24($-$09) & 1.11($-$09) & 2.15($-$10) \\
3s ($^2$S$^\mathrm{e}$) &    & - -   &    9.63($-$07) & 7.18($-$07) & 4.21($-$08) & 3.76($-$08) & 3.70($-$08) & 9.52($-$08) & 1.46($-$08) \\
3p ($^2$P$^\mathrm{o}$) &    &    &  - -  &    1.86($-$05) & 1.43($-$07) & 1.42($-$07) & 1.57($-$07) & 2.75($-$07) & 1.84($-$08) \\
3d ($^2$D$^\mathrm{e}$) &    &    &    & - -   &    9.27($-$08) & 8.23($-$08) & 1.72($-$07) & 3.30($-$07) & 1.57($-$08) \\
4s ($^2$S$^\mathrm{e}$) &    &    &    &    & - -   &    4.80($-$06) & 3.37($-$06) & 1.73($-$06) & 2.43($-$07) \\
4p ($^2$P$^\mathrm{o}$) &    &    &    &    &    & - -   &    4.80($-$05) & 9.41($-$06) & 7.60($-$07) \\
4d ($^2$D$^\mathrm{e}$) &    &    &    &    &    &    & - -   &    3.90($-$05) & 4.05($-$07) \\
4f ($^2$F$^\mathrm{o}$) &    &    &    &    &    &    &    &  - -  &    2.13($-$07) \\

  &  \multicolumn{9}{c}{\bf 12 000 K}  \\
2s ($^2$S$^\mathrm{e}$) &    8.67($-$08) & 2.15($-$09) & 9.95($-$10) & 1.29($-$09) & 2.54($-$10) & 1.74($-$10) & 2.75($-$10) & 1.73($-$10) & 1.12($-$10) \\
2p ($^2$P$^\mathrm{o}$) &  - -  &    2.46($-$08) & 1.92($-$08) & 3.36($-$08) & 1.58($-$09) & 3.06($-$09) & 4.98($-$09) & 4.19($-$09) & 7.83($-$10) \\
3s ($^2$S$^\mathrm{e}$) &    & - -   &    1.56($-$06) & 8.72($-$07) & 5.99($-$08) & 6.46($-$08) & 6.32($-$08) & 1.81($-$07) & 2.62($-$08) \\
3p ($^2$P$^\mathrm{o}$) &    &    &  - -  &    1.68($-$05) & 2.01($-$07) & 2.04($-$07) & 2.44($-$07) & 4.56($-$07) & 2.86($-$08) \\
3d ($^2$D$^\mathrm{e}$) &    &    &    & - -   &    1.01($-$07) & 9.88($-$08) & 2.49($-$07) & 5.94($-$07) & 2.02($-$08) \\
4s ($^2$S$^\mathrm{e}$) &    &    &    &    & - -   &    7.10($-$06) & 4.04($-$06) & 1.69($-$06) & 3.08($-$07) \\
4p ($^2$P$^\mathrm{o}$) &    &    &    &    &    & - -   &    4.19($-$05) & 7.77($-$06) & 9.94($-$07) \\
4d ($^2$D$^\mathrm{e}$) &    &    &    &    &    &    & - -   &    3.24($-$05) & 4.12($-$07) \\
4f ($^2$F$^\mathrm{o}$) &    &    &    &    &    &    &    &  - -  &    1.90($-$07) \\

    & \multicolumn{9}{c}{\bf 20 000 K}  \\
2s ($^2$S$^\mathrm{e}$) &    1.98($-$07) & 6.10($-$09) & 3.74($-$09) & 5.91($-$09) & 1.18($-$09) & 9.68($-$10) & 1.78($-$09) & 9.36($-$10) & 5.96($-$10) \\
2p ($^2$P$^\mathrm{o}$) &  - -  &    3.79($-$08) & 3.65($-$08) & 7.77($-$08) & 3.74($-$09) & 8.49($-$09) & 1.62($-$08) & 1.20($-$08) & 2.00($-$09) \\
3s ($^2$S$^\mathrm{e}$) &    & - -   &    2.53($-$06) & 9.79($-$07) & 8.74($-$08) & 1.03($-$07) & 1.03($-$07) & 2.99($-$07) & 4.06($-$08) \\
3p ($^2$P$^\mathrm{o}$) &    &    &  - -  &    1.39($-$05) & 2.91($-$07) & 2.94($-$07) & 3.85($-$07) & 7.01($-$07) & 4.07($-$08) \\
3d ($^2$D$^\mathrm{e}$) &    &    &    & - -   &    1.01($-$07) & 1.11($-$07) & 3.40($-$07) & 1.06($-$06) & 2.18($-$08) \\
4s ($^2$S$^\mathrm{e}$) &    &    &    &    & - -   &    1.01($-$05) & 4.59($-$06) & 1.56($-$06) & 4.02($-$07) \\
4p ($^2$P$^\mathrm{o}$) &    &    &    &    &    & - -   &    3.36($-$05) & 5.95($-$06) & 1.24($-$06) \\
4d ($^2$D$^\mathrm{e}$) &    &    &    &    &    &    & - -   &    2.50($-$05) & 3.90($-$07) \\
4f ($^2$F$^\mathrm{o}$) &    &    &    &    &    &    &    &  - -  &    1.52($-$07) \\

   & \multicolumn{9}{c}{\bf 50 000 K}  \\
2s ($^2$S$^\mathrm{e}$) &    4.59($-$07) & 1.60($-$08) & 1.23($-$08) & 2.55($-$08) & 4.36($-$09) & 4.11($-$09) & 9.07($-$09) & 3.49($-$09) & 2.10($-$09)\\
2p ($^2$P$^\mathrm{o}$) &  - -  &    6.40($-$08) & 6.89($-$08) & 2.08($-$07) & 7.82($-$09) & 2.03($-$08) & 4.91($-$08) & 2.74($-$08) & 3.73($-$09) \\
3s ($^2$S$^\mathrm{e}$) &    & - -   &    4.25($-$06) & 1.00($-$06) & 1.43($-$07) & 1.53($-$07) & 1.61($-$07) & 3.95($-$07) & 5.35($-$08) \\
3p ($^2$P$^\mathrm{o}$) &    &    &  - -  &    8.96($-$06) & 4.60($-$07) & 4.34($-$07) & 6.67($-$07) & 9.68($-$07) & 5.58($-$08) \\
3d ($^2$D$^\mathrm{e}$) &    &    &    & - -   &    8.45($-$08) & 1.08($-$07) & 4.28($-$07) & 1.97($-$06) & 1.74($-$08) \\
4s ($^2$S$^\mathrm{e}$) &    &    &    &    & - -   &    1.28($-$05) & 4.57($-$06) & 1.24($-$06) & 5.48($-$07) \\
4p ($^2$P$^\mathrm{o}$) &    &    &    &    &    & - -   &    2.10($-$05) & 3.54($-$06) & 1.37($-$06) \\
4d ($^2$D$^\mathrm{e}$) &    &    &    &    &    &    & - -   &    1.50($-$05) & 3.11($-$07) \\
4f ($^2$F$^\mathrm{o}$) &    &    &    &    &    &    &    & - -   &    8.78($-$08) \\
\hline \hline
	\end{tabular}
	\tablefoot{Rate coefficients, $\langle\sigma v\rangle$, are given in cm$^3$ s$^{-1}$.  The values should be read as: $a(b) := a \times 10^b$.}
	}
\end{table*}

The non-LTE line source function for the very important 6708~{\AA} resonance line is described well by a two-level approximation \citep{1994A&A...288..860C}, and thus the data for $2s \rightarrow 2p$ are of special interest.  The data are compared in Table~\ref{tab:2s2p} and all four calculations are in reasonable agreement, though the \citet{Park1971} value is roughly a factor of two larger than the other three.

\begin{table}
\caption{Comparison of the rate coefficient $\langle \sigma v \rangle$ for the transition $2s \rightarrow 2p$ at 8000 K.}\label{tab:2s2p}
\centering
 \begin{tabular}{cc}
 \hline
 Method   & $\langle \sigma v \rangle$ (cm$^3$ s$^{-1}$) \\\hline\hline
    This Work           &    3.28($-08$)   \\
   \cite{CCC}          &   3.33($-08$)   \\
   \cite{Park1971}  &  6.36($-08$)   \\
   \cite{1962ApJ...136..906V}      &  2.46($-08$)   \\
   \hline \hline
 \end{tabular}
\end{table}

\subsection{Ionization}

\begin{figure*}[t]
\centering
	\begin{tabular}{c c c}
\includegraphics[width=0.95\textwidth]{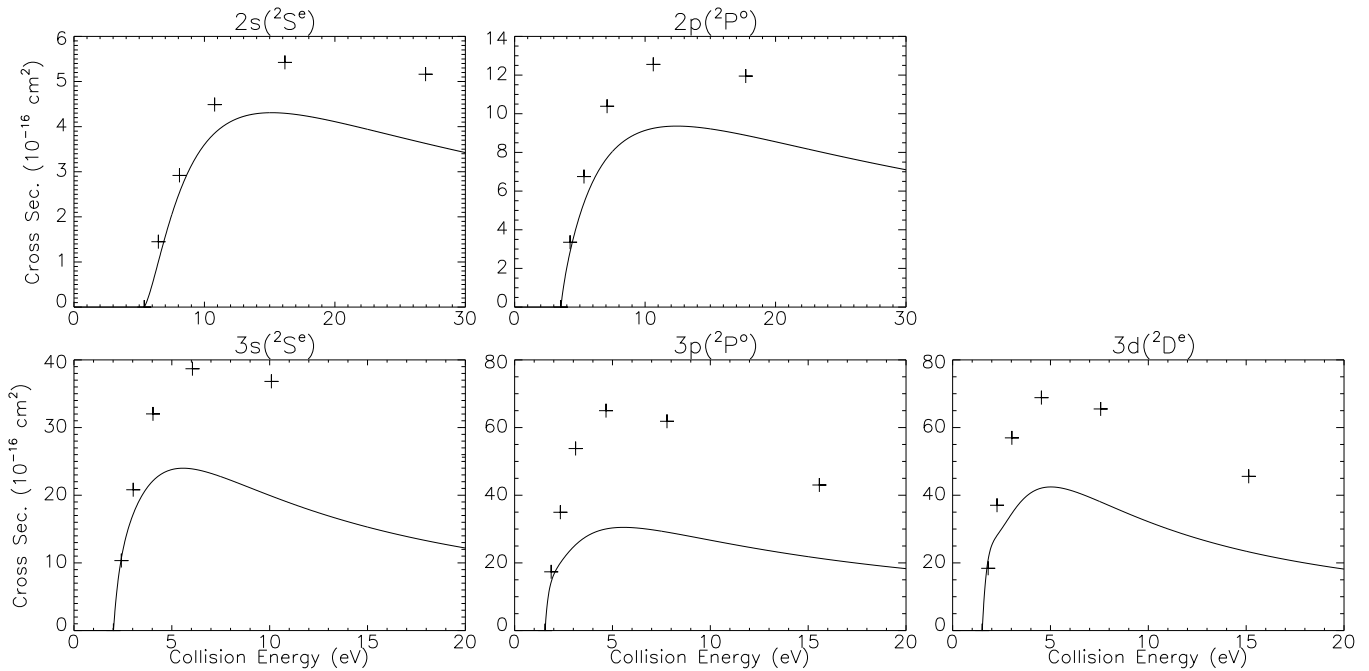}
	\end{tabular}
     \caption{Comparison of electron-impact ionization cross sections with the collision energy for the lowest states of Li I. The solid lines show cross sections from \cite{lennon:1285} and the CCC calculations \citep{CCC} (see text).  The plus signs are those calculated from the semi-empirical formula \citep{1976asqu.book.....A}.}\label{ioncrossec}
\end{figure*}

Regarding electron-impact ionization $\mathrm{Li}(nl) + e \rightarrow \mathrm{Li}^+ + 2e$, \citet{CCC} have also calculated cross sections via the CCC method for the $2p$ and $n = 3$ states. For ionization cross sections from the ground state ($2s$) calculations by \cite{lennon:1285} are recommended by \cite{CCC}.  The results of CCC and RMPS calculations for this process where $n=2$ have been shown to be in good agreement \citep{2001PhRvL..87u3201C}.  Thus, since this process is likely to be less important in this application, we adopted the CCC data directly.  As mentioned, non-LTE calculations have typically used a semi-empirical formula from \citet{1976asqu.book.....A} based on a fit to experimental data \citep[see][]{1962amp..conf..375S}.   A comparison of cross sections from the semi-empirical formula with those of \citet{CCC} is presented in Fig.~\ref{ioncrossec}.   The comparison shows a tendency for the ionization cross sections of the semi-empirical formula to be significantly  larger than the CCC results.   Rate coefficients calculated by integration over a Maxwellian velocity distribution are presented in Table~\ref{tab:ion}.

\begin{table*}
\caption{Rate coefficients for collisional ionization between electrons and Li I atoms for different temperatures. }
\label{tab:ion}
\centering
	\begin{tabular}{l c c c c c c c}\hline
Bound & \multicolumn{7}{c}{Temperature (K)} \\
State     &   1 000  &    2 000 &      5 000 &      8 000 &      12 000 &      20 000 &      50 000 \\\hline\hline\\
2s($^2$S$^\mathrm{e}$) &   4.85($-$36) & 3.00($-$22) & 8.63($-$14) & 1.35($-$11) &  2.46($-$10) &  2.73($-$09)  & 2.69($-$08) \\[0.1cm]
2p($^2$P$^\mathrm{o}$) & 5.53($-$26)  & 6.19($-$17) & 2.06($-$11)  & 5.49($-$10)  & 3.62($-$09)  & 1.76($-$08) & 8.23($-$08) \\[0.1cm]
3s($^2$S$^\mathrm{e}$) & 1.03($-$17) & 1.57($-$12) & 2.34($-$09)  & 1.55($-$08)  & 4.55($-$08)  & 1.11($-$07) & 2.44($-$07) \\[0.1cm]
3p($^2$P$^\mathrm{o}$) & 3.10($-$15) & 3.03($-$11) & 8.07($-$09) & 3.45($-$08) & 8.07($-$08)  & 1.67($-$07) &  3.36($-$07) \\[0.1cm]
3d($^2$D$^\mathrm{e}$) &  7.00($-$15) & 5.43($-$11) & 1.25($-$08) & 5.11($-$08) & 1.16($-$07) & 2.33($-$07) & 4.30($-$07) \\ \\\hline\hline
	\end{tabular}
\tablefoot{The rate coefficients $\langle\sigma v\rangle$, in cm$^3$ s$^{-1}$, are calculated from the ionization cross sections from \cite{lennon:1285} and \cite{CCC} (see text).  The values should be read as: $a(b) := a \times 10^b$}
\end{table*}

\section{Application to non-LTE Li I line formation }

The most commonly observed lines for Li I in stellar spectra are those corresponding to the $2s$-$2p$ and the $2p$-$3d$ transitions at 6708 \AA\ and 6104 \AA , respectively. In Li-rich stars the 8126~\AA\ line is also observed \citep{8126lineA,8126lineB}, this line corresponding to the transition $2p$-$3s$.  It has also been suggested that this line might be visible in other Li-rich objects such as chemically peculiar stars \citep{1999A&A...351..283P}, though we are not aware of any such detections.

\begin{figure*}
\centering
	\begin{tabular}{l c r}
	\includegraphics[width=0.33\textwidth]{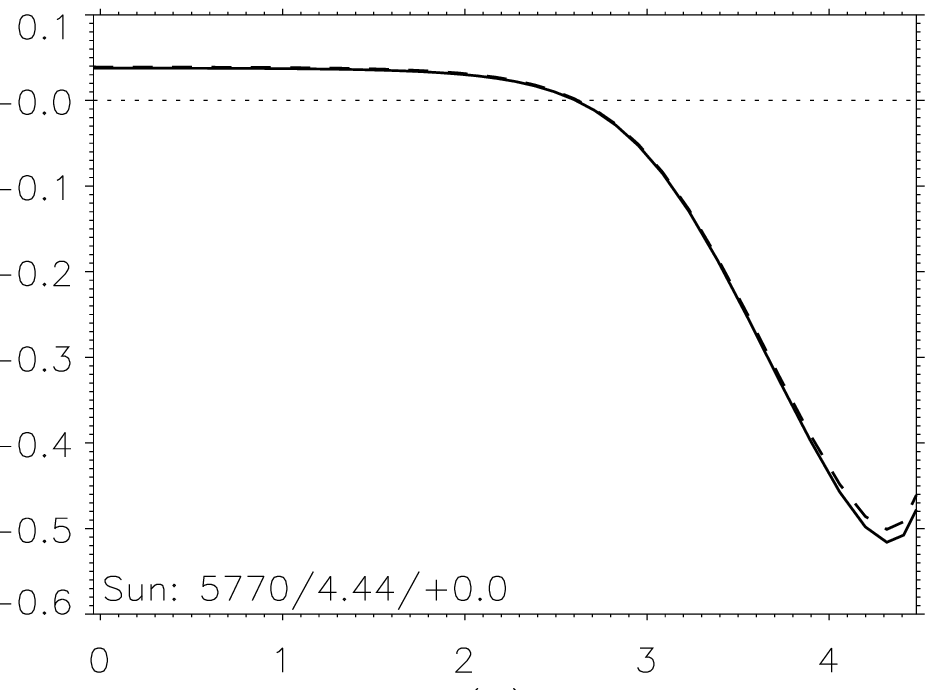}&  &  \includegraphics[width=0.33\textwidth]{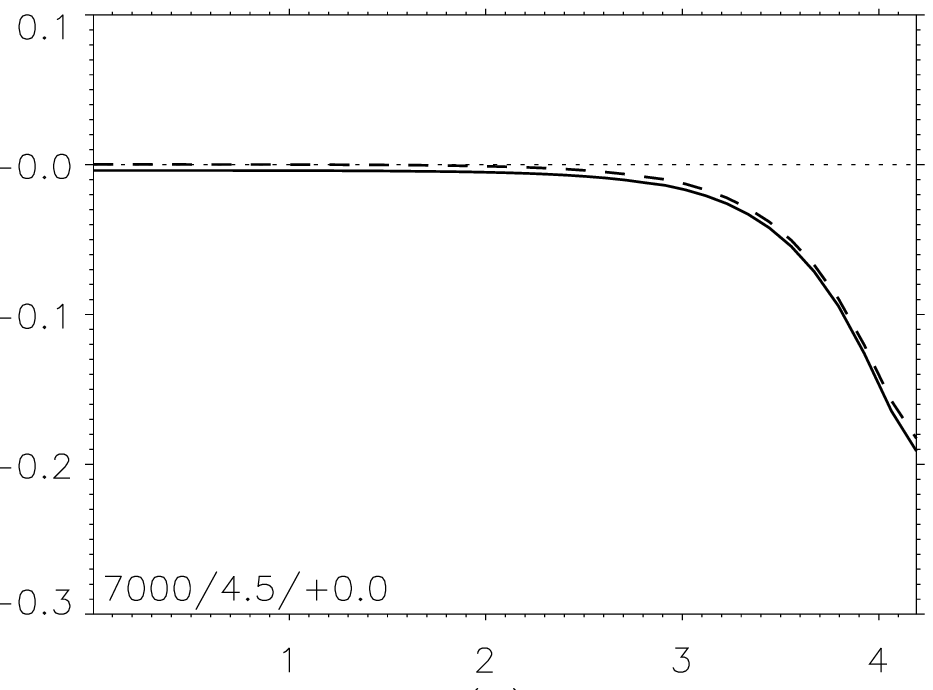} \\
	&&\\
	\includegraphics[width=0.33\textwidth]{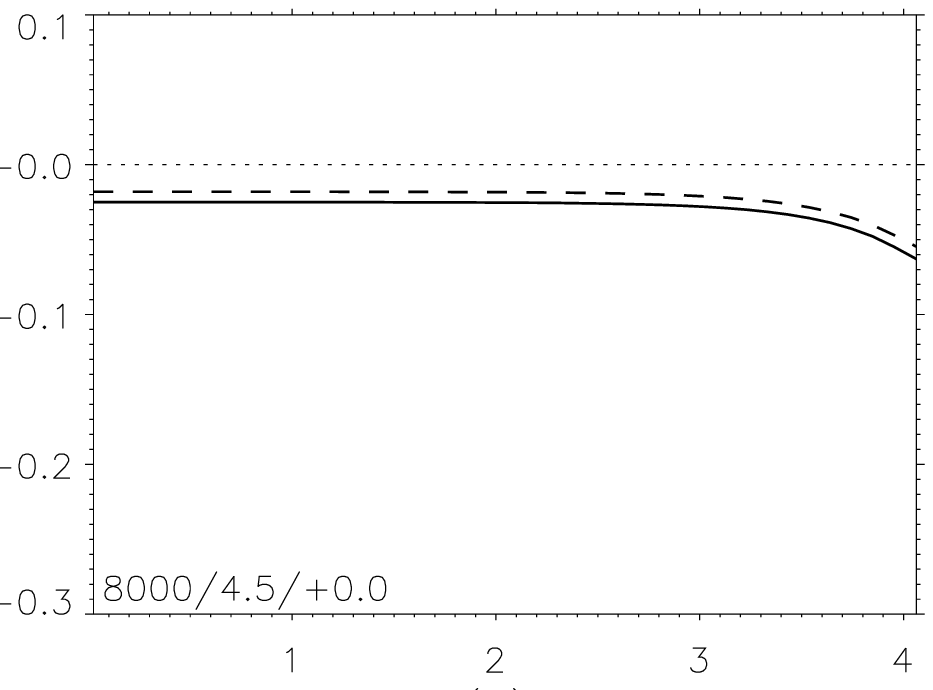}& & \includegraphics[width=0.33\textwidth]{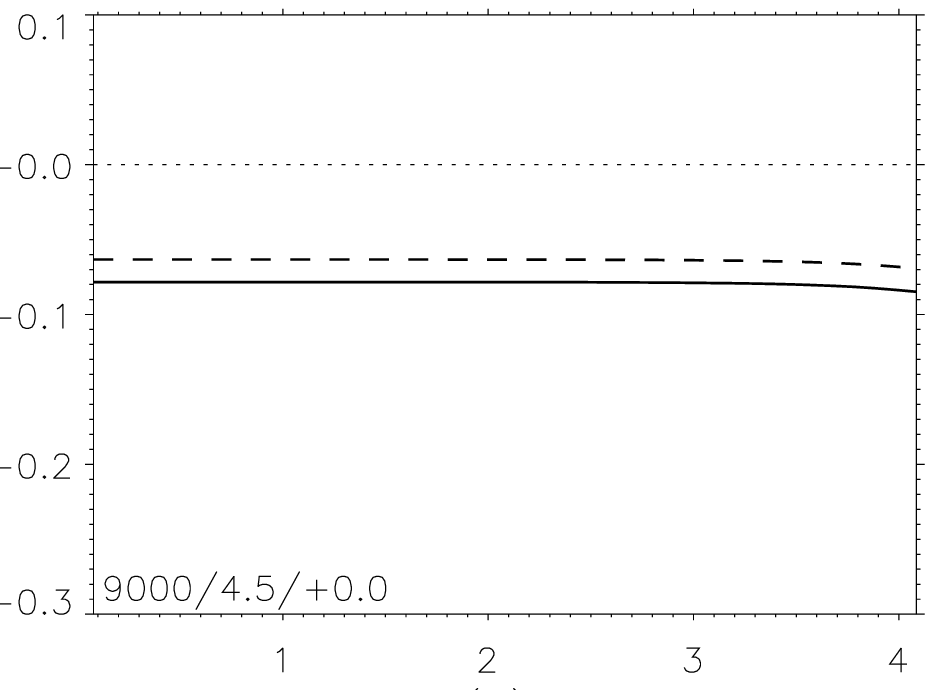}
	\end{tabular}
\caption{Non-LTE Li abundance corrections for the 6707~\AA\ resonance line using the model atom as \cite{Lind2009} and replacing the e+Li collisions from \cite{Park1971} (dashed line) with the new calculations (solid line). The dotted line shows the zero value. In the lower left corner of each plot are the stellar parameters of the model used: \mbox{$T_\mathrm{eff}$/log $g$/[Fe/H]}.}\label{acorr}
\end{figure*}

Previous non-LTE calculations of Li line formation in cool stars by \cite{1994A&A...288..860C} and \citet{Lind2009} used the rate coefficients of \cite{Park1971} for the description of the electron-impact excitation in their model atom and the recipe of \citet{1976asqu.book.....A} for ionization.  \citet{1984A&A...130..319S} also used the Park data, but replaced the value for the $2s$-$2p$ transition with the van Regemorter value.  A similar semi-empirical recipe for the ionization is also used.  Thus, the three works that are of most interest for comparison use essentially the same electron collision data.   In this work we used the same model atom as \citet{Lind2009}, changing only the electronic collision data; that is, the rate coefficients of Tables~\ref{tablerates} and \ref{tab:ion} replace the old data for excitation and ionization by electron impact.  The collision rates from \cite{Park1971} for the transitions involving high-lying levels not covered by our calculations were retained.   We calculated non-LTE corrections for a range of stellar atmospheric models using the old and new model atoms.  Some example comparisons of the results are shown in Fig.~\ref{acorr}.   We found the differences between the two model atoms to be $<0.01$ dex for solar-type stars. In warm stars, $T_\mathrm{eff}>7000$~K, the differences are slightly larger but still $\la 0.02$\,dex.  

The behaviour of the abundances measured using the 6104~\AA\ and  8126~\AA\ spectral lines was also studied. The transition  $2p$-$3s$, corresponding to the 8126~\AA\ line, has the largest difference in the rate coefficients when compared with those from \citet{Park1971}.  For all the atmospheric model cases shown in Fig.~\ref{acorr}, the Li abundance derived from this line shows very similar sensitivity to the new electron collision data.  The effects of the new electron collision data on the abundance measured from the 6104~\AA\ line are considerably less in the solar case: only about 10\% of the difference shown by the other two lines.  In the other cases the influence of the electron collision data on the abundance measured from the 6104~\AA\ spectral line is again comparable to the influence on the 6708 and 8126~\AA\ lines.

\section{Conclusions}
A common problem for calculating non-LTE abundance corrections is that they can be subject to errors from uncertainties in the underlying atomic data.   We have shown that electron collision data from a number of different calculations based on quite different methods are in good or even excellent agreement.  Data calculated with advanced close-coupling techniques provides data in excellent agreement, giving rate coefficients for excitation processes within a factor of 2 for the temperatures of most interest in cool stellar atmospheres, $T \sim 5000$ -- 8000~K.  This, and the good agreement with experimental results where available \citep[see][]{CCC}, suggests the data have an uncertainty of similar magnitude.   The semi-empirical data for both excitation and ionization that have been commonly used in non-LTE calculations for Li up to now, agree reasonably with the modern data, generally agreeing within a factor of 6, though with a few significant outliers.  This suggests that the estimates of the uncertainties of a factor of 2 for these data are somewhat too low, though the data for the important $2s$-$2p$ transition does indeed agree within a factor a two.  

We find that these uncertainties are not very important when applied to Li I in cool stars, and the differences typically result in uncertainties of less than 0.01 dex.  The strongest effects are found in F dwarfs and extremely Li-rich stars.  These uncertainties are certainly negligible compared to those arising from other sources, especially the atmospheric modelling \citep{2005ARA&A..43..481A}.   
Thus, we conclude that the existing electron collision data, or the data presented in this paper, when coupled with data for the charge exchange process $\mathrm{Li(3s)} + \mathrm{H} \rightleftharpoons \mathrm{Li}^+ + \mathrm{H}^-$ now available, are accurate enough to reliably model non-LTE formation of Li lines in stellar atmospheres at the 0.01~dex ($\sim 2$\%) level.

\begin{acknowledgements} 
We thank Martin Plummer for providing us with an optimised copy of the \texttt{CIVPOL} code.  We gratefully acknowledge the support of the Royal Swedish Academy of Sciences, G{\"o}ran Gustafssons Stiftelse, and the Swedish Research Council.  Y.O.\ is grateful for hospitality and financial support during a visit to the Max Planck Institute for Astrophysics, Garching. P.S.B.\ is a Royal Swedish Academy of Sciences Research Fellow supported by a grant from the Knut and Alice Wallenberg Foundation.   
\end{acknowledgements}

\bibliographystyle{aa}       		
\bibliography{16418}		
\end{document}